\documentclass[12pt,a4paper]{article}
\usepackage{graphicx,amssymb}
\usepackage{graphicx, array, epsfig, subfigure}
\usepackage{epstopdf}
\usepackage{subfigure}  
\newcommand{\be}{\begin{equation}}
\newcommand{\ee}{\end{equation}}
\newcommand{\bea}{\begin{eqnarray}}
\newcommand{\eea}{\end{eqnarray}}

\catcode`@=12

\begin{document}
\begin{center}
{\bf Unparticle Decay of Neutrinos and its Possible Signatures at 
a ${\rm Km}^2$ Detector for (3+1) Flavour Framework}
\end{center}
\vspace{1cm}
\begin{center}
{ {\bf Madhurima Pandey} \footnote {email: madhurima.pandey@saha.ac.in}}
\end{center}
\vskip 0.5mm
\begin{center}
Astroparticle Physics and Cosmology Division  \\
Saha Institute of Nuclear Physics, HBNI  \\
1/AF Bidhannagar, Kolkata 700064, India  \\
\end{center}
\vskip 3mm
\begin{center}
{\bf Abstract}
\end{center}
{\small
We consider a scenario where ultra high energy neutrinos undergo unparticle 
decay during its passage from its cosmological source to Earth. 
The idea of unparticle had been first proposed by Georgi by considering the 
possible existence of an unknown scale invariant sector at high energies 
and the unparticles in this sector manifest itself below a dimensional 
transmutation scale $\Lambda_{\cal U}$. We then explore the possible signature 
of such decaying neutrinos to unparticles at a square kilometer detector 
such as IceCube.}
\newpage
\section{Introduction}
Almost a decade back Georgi \cite{georgi1,georgi2} 
proposed the probable existence of a scale 
invariant sector. At a very high energy scale this scale invariance sector 
and the Standard Model (SM) sector may coexist and the fields of these two 
sectors can interact via a mediator messenger field of mass scale $M_{\cal U}$.
This is the connector sector \cite{kingman}. At low energies however, the 
scale invariance is manifestly broken since SM particles have masses.
At a scale below $M_{\cal U}$ such interactions are suppressed by inverse 
powers 
of $M_{\cal U}$ and the effective theory at low energy can be expressed 
by a non-renormalizable operator. It is also to be noted that in a scale 
invariant scenario the particle masses are zero and in the real world, 
the scale invariance is manifestly broken. It is observed by Georgi 
\cite{georgi1,georgi2} that at low  energies such a scale invariance sector 
of scale dimension $d_{\cal U}$ manifests itself as non-integral 
number $d_{\cal U}$ of massless invisible particles called ``unparticles".   

It is to be noted that in 4-D Quantum Field Theory (QFT), the conformal 
invariance is broken by 
renormalization group effects. but such a conformal invariance in 
4-D can be described by a vector like non 
Abelian gauge theory studied by Banks and Zaks (BZ) \cite{bankszaks}.
In this theory the scale invariant sector can flow to low energies 
with nontrivial infrared fixed points and the theory 
may be extended to low energy. Following Georgi's proposal, the 
interaction operator 
${\cal O}_{\rm BZ}$ for the BZ fields with 
the operator ${\cal O}_{\rm SM}$ for SM fields can generically 
be represented by  
${\cal O}_{\rm BZ}{\cal O}_{\rm SM}/(M_{\cal U}^k), k > 0$. In a massless
non abelian gauge theory, the radiative corrections in the scale invariant
sector induce dimensional transmutation \cite{weinberg} at another
energy scale. As a result, 
another scale $\Lambda_{\cal U}$ appears and Georgi argued 
\cite{georgi1,georgi2} that below this scale the BZ field and field 
operator ${\cal O}_{\rm BZ}$ matches onto the unparticle operator
${\cal O}_{\cal U}$ with non-integral scaling dimension $d_{\cal U}$.
Thus below $\Lambda_{\cal U}$, one has new low energy operator 
of the form $C_{{\cal O}_{\cal U}}\Lambda_{\cal U}^{d_{BZ} - d_{\cal U}}
{\cal O}_{\rm SM}{\cal O}_{\cal U}/(M_{\cal U}^k)$, where 
$C_{{\cal O}_{\cal U}}$ is to be fixed from the matching conditions 
of BZ operator ${\cal O}_{\rm BZ}$ onto the unparticle operator 
${\cal O}_{\cal U}$. In this operator $d_{\rm BZ}$ denotes the scaling 
dimension 
of the operator ${\cal O}_{\rm BZ}$. Since at low energies ${\rm BZ}$ fields 
decouple 
from the SM fields. The infrared fixed ponts of the unparticles will remain 
unaffected by the couplings of the unparticle and the SM particles.

The unparticle physics gives rise to rich phenemenology of many unexpected 
processes. Several authors in the literatures used the concept of unparticles 
in a wide range of particle physics issues. For example Kikuchi and Okada 
\cite{kikuchi} addressed the unparticle couplings with Higgs and gauge bosons. 
The interactions of unpartilces with SM particles are addressed by various 
other authors \cite{authors}. The issues of dark matter and dark energy is 
discussed in the unparticle framework in the works of Refs. \cite{dm}. We 
consider the unparticle decay of neutrinos and explore its consequences for 
Ultra High Energy (UHE) neutrinos from a distant Gamma Ray Bursts (GRBs). 
For this case, the decay length should be $\sim$ tens of Mpc for such decay 
is to be significant. Here we investigate the unparticle decay of neutrinos 
along with the mass flavour suppression due to passage of such UHE neutrinos 
from a distant GRB to an Earth ground detector such as IceCube. We also 
consider a four flavour scenario for the neutrino species where we assume a 4th sterile species along with the usual 3 active neutrinos. The possible existence of the sterile neutrino as already been indicated by the neutrino 
experiments such as MINOS \cite{minos}-\cite{Adamson:2016jku}, 
Daya Bay \cite{Adamson:2016jku}-\cite{daya6}, Bugey \cite{bugey} etc. We 
calculate the neutrino induced 
muon yield in such a scenario at a square kilometer detector 
such as IceCube.    

The paper is organised as follows. A brief account of the formalism of UHE 
neutrinos, which decay to unparticle decay from a single GRB is discussed in 
Section 2. We have considered three active and one sterile neutrinos 
(3+1) framework in the present work. Section 2 is divided into two subsections. 
In Subsection 2.1 we address the expression for the neutrino spectrum on 
reaching the Earth from a single GRB in the absence of decay or oscillations, 
while the form of the UHE neutrino fluxes, considering the unparticle 
decay phenomenon, from a single GRB at redshift $z$ is furnished in 
Subsubsection 2.1.1. In Subsection 2.2 we describe the analytical expressions 
for the total number of neutrino induced muons from a point like source 
such as a single GRB at a square kilometer detector such as IceCube. The 
calculational results of the yield of secondary muons in different scenarios 
are given in Section 3. Finally in Section 4 we give a brief summary and 
discussions.
\section{Formalism}
\subsection{UHE neutrino fluxes from a single GRB with neutrino decay to 
unparticles}
Gamma-Ray Bursts (GRBs) \cite{wax} are some of the most energetic events in the 
Universe. We have considered the relativistically expanding fireball model, 
which is one of the few models that has been put forth to explain why GRBs 
tend to have such high energy levels. In this model, the Fermi mechanism 
in shocks developing in the GRB outflow can accelerate protons to energies 
as high as $10^{20}$ eV. These highly energetic accelerated protons interact 
with photons via a cosmic beam dump process inside the fireball and the pions 
are produced through these interactions. In our work we consider the UHE 
neutrinos which are produced by the decay of these pions and the decay 
process is $\pi^{+} \rightarrow \mu^{+} + \nu_{\mu}$, which is followed by 
the muons decaying to $\mu^{+} \rightarrow e^{+} + \nu_e + \bar{\nu}_{\mu}$.

There are some parameters, which are required to calculate the GRB neutrino 
spectrum, like Lorentz factor ${\Gamma}$ ($\Gamma$ plays an 
important role in the neutrino production mechanism of the GRB), neutrino 
break energy $E_{\nu}^{\rm brk}$, observed photon spectral break energy 
$E_{\gamma, {\rm MeV}}^{\rm brk}$, the total amount of energy released at the time of neutrino emission $E_{\rm GRB}$ ($E_{\rm GRB} = 10^{53} {\rm erg}$, which 
is 10$\%$ of the fireball photon energy), the wind variability time $t_{\nu}$, 
redshift distance of GRB from the observer ($z$) and the wind luminosity 
$L_w$ ($\simeq 10^{53}$ erg/sec) \cite{nayantara,guetta}.  

The neutrino spectrum of the GRB \cite{nayantara,guetta,nayantara1} can be 
written as
\bea
\displaystyle\frac {dN_{\nu}} {dE_{\nu}} = N \times {\rm min}
\left (1,\displaystyle\frac {E_{\nu}} {E_{\nu}^{\rm brk}} \right ) \displaystyle\frac {1} {E_{\nu}^2}\,\, .
\label{det1}
\eea
In the above, $N$ represents the normalization constant and $E_{\nu}$ is the
neutrino energy. The neutrino apectrum break energy $E_{\nu}^{\rm brk}$ can be 
expressed in terms of the Lorentz boost factor ($\Gamma$) and the photon 
spectral break energy ($E_{\gamma, {\rm MeV}}^{\rm brk}$).
\bea
E_{\nu}^{\rm brk} \approx 10^6 \displaystyle\frac {\Gamma_{2.5}^2}
{E_{\gamma , \rm MeV}^{\rm brk}} \rm GeV\,\, ,
\label{det2}
\eea
where $\Gamma_{2.5} = \Gamma/10^{2.5}$. The normalization constant ($N$), 
which is mentioned in Eq. (\ref{det1}), is given by
\bea
N = \displaystyle\frac {E_{\rm GRB}}
{1 + \rm ln (E_{\nu \rm max} / E_{\nu}^ {\rm brk})}\,\, .
\label{det3}
\eea
The lower and the upper cut-off energy of the neutrino spectrum are denoted by 
$E_{\nu{\rm min}}$ and $E_{\nu{\rm max}}$ respectively.

At a particular distance of the GRB from the observer ($z$), the relation 
between the observed neutrino energy $E_{\nu}^{\rm obs}$ and the actual 
energy of neutrino at the source $E_{\nu}$ is given as
\bea 
E_{\nu}^{\rm obs} &=& \displaystyle \frac {E_{\nu}}{(1 + z)}\,\, .
\label{det4}
\eea
Likewise for the upper cut-off energy of the source Eq. (\ref{det4}) can be 
written as
\bea 
E_{\nu {\rm max}}^{\rm obs} &=& \displaystyle \frac {E_{\nu {\rm max}}}{(1 + z)}
\,\, .
\label{det5}
\eea

Thus in the absence of decay or oscillation the neutrino spectrum on reaching 
the Earth from a GRB at redshift $z$ takes the form.
\bea
\displaystyle\frac {dN_{\nu}}{dE_{\nu}^{ \rm obs}} = 
\displaystyle\frac
{dN_{\nu}} {dE_{\nu}} \displaystyle\frac {1} {4\pi r^2(z)} (1 + z)\,\, .
\label{det6}
\eea

In the absence of CP violation ${\cal F} (E_\nu) = \displaystyle\frac {dN_\nu}
{dE_\nu^{\rm obs}} = \displaystyle\frac {dN_{\nu + \bar{\nu}}} 
{dE_\nu^{\rm obs}}$. The spectra 
for neutrinos will be 0.5${\cal F}(E_\nu)$. 

Now the neutrinos are produced in the GRB process in the proportion
\bea
\nu_e:\nu_\mu:\nu_\tau:\nu_s &=& 1:2:0:0\,\, .
\label{ratio}
\eea
Therefore 
\bea
\phi_{\nu_e}^s  =  \displaystyle \frac{1} {6} {\cal F}(E_\nu)\,\, ,
\phi_{\nu_\mu}^s  =  \displaystyle \frac{2} {6} {\cal F}(E_\nu)  =  2\phi_{\nu_e}^s\,\, ,
\phi_{\nu_\tau}^s = 0\,\, ,
\phi_{\nu_s}^s = 0\,\, ,
\label{flux}
\eea
where $\phi_{\nu_e}^s$, $ \phi_{\nu_\mu}^s$, $\phi_{\nu_\tau}^s$ and 
$\phi_{\nu_s}^s$ are the fluxes of $\nu_e$, $\nu_\mu$, $\nu_\tau$ and 
$\nu_s$ at source repectively.

In Eq. (\ref{det6}) $r(z)$ denotes the comoving radial coordinate distance of 
the source, which can be expressed as
\bea
r(z) = \displaystyle\frac {c} {H_0} \int_{0}^{z} \displaystyle\frac {dz'}
{\sqrt{\Omega_{\Lambda} + \Omega_{m} (1 + z')^3}}\,\, .
\label{det7}
\eea
$\Omega_{\Lambda} + \Omega_m = 1$ for spatially flat Universe, where 
$\Omega_{\Lambda}$ is the contribution of dark energy density in units of the 
critical energy density of the Universe and $\Omega_m$ represents the 
contribution of the matter to the energy density of the Universe in units of 
critical density.

In Eq. (\ref{det7}), $c$ and $H_0$ denote respectively the speed of the light 
and the Hubble constant in the present epoch. The values of the constants 
which we have used in our calculations are $\Omega_\Lambda = 0.68$,  
$\Omega_m = 0.3$ and $H_0 = 73.8$ $\rm Km\,\, \rm sec^{-1}\,\, \rm Mpc^{-1}$.
\subsubsection{Unparticle decay of GRB neutrinos}
After the Georgi's ``Unparticle" proposal, 
extensive studies to investigate the 
unparticle phenomenology
have been explored in the literature. Unparticle physics 
is a speculative theory that conjectures a form of matter that cannot be 
explained in terms of particles using the Standard Model (SM) of particle 
physics, because its components are scale invariant. So the interaction 
between the unparticle and SM particles is speculative in nature. The presence 
of this unparticle operator can effect the processes, which are all measured 
in experiments. Some processes where the invisible unparticles (${\cal U}$) has 
been considered as the final state are (1) the top quark decay $ \tau 
\rightarrow u + {\cal U}$ \cite{georgi1}, (2) the electron - positron annihilation $ e^{+} + 
e^{-} \rightarrow \gamma +{\cal U}$ , (3) the hadronic processes such as 
$q + q \rightarrow g + {\cal U}$ \cite{georgi2,kingman} etc.

In the present work we consider a decay phenomenon , where neutrino having mass eigenstate $\nu_{j}$ decays to the invisible unparticle (${\cal U}$) 
\cite{shun} and another light neutrino with mass eigenstate $\nu_{i}$.
\be
\nu_{j} \rightarrow {\cal U} + \nu_{i}\,\, .
\label{form1}
\ee

The effective lagrangian for the above mentioned process takes the following 
form in the low energy regime.
\be
L_{s} = \displaystyle\frac {\lambda_{\nu}^{\alpha \beta}} {\Lambda_{\cal U}^{d_{\cal U} -1}} \bar{\nu}_{\alpha} \nu_{\beta} {\cal O}_{\cal U} \,\, ,
\label{form2}
\ee
where $\alpha, \beta = e, \mu, \tau, s$ are the flavour indices, 
$d_{\cal U}$ is the 
scaling dimension of the scalar unpartcile operator ${\cal O}_{\cal U}$. 
$\Lambda_{\cal U}$ and $\lambda_\nu^{\alpha \beta}$ indicate the dimension 
transmutation 
scale at which the scale invariance sets in and the relevant coupling constant 
respectively. From Eq. (\ref{form2}). note that a heavier neutrino decays into 
a lighter neutrino and an unparticle.

The neutrino mass and flavour eigenstates are related through
\bea
|\nu_i \rangle &=& \displaystyle\sum_{\alpha} U_{\alpha i}^{*} |\nu_{\alpha}  
\rangle\,\, ,
\label{form3}
\eea
where $U_{\alpha i}$ are the elements of the Pontecorvo - Maki - Nakagawa - 
Sakata (PMNS) \cite{maki} mixing matrix. Working in the neutrino mass eigen state basis is 
more convenient than the flavour eigenstate. So in this mass basis we can write 
the interaction term bettween neutrinos and the unparticles as 
$\lambda_\nu^{ij} \bar{\nu}_i \nu_j {\cal O}_{\cal U}/\Lambda_{\cal U}^{d_{\cal U} - 1}$ , where $\lambda_\nu^{ij}$ is the coupling constant in the mass eigenstate $i,j$.

Now the above mentioned coupling constant can be expressed as 
\bea
\lambda_\nu^{ij} &=& \displaystyle\sum_{\alpha, \beta} U_{\alpha i}^{*} 
\lambda_\nu^{\alpha \beta} U_{\beta j}\,\, .
\label{form4} 
\eea

The total decay rate $\Gamma_{j}$ or equivalently the lifetime of neutrino 
$\tau_{\cal U} = 1/{\Gamma_{j}}$ is the most relevant quantity for the decay 
process   $\nu_{j} \rightarrow {\cal U} + \nu_{i}$ \cite{shun}. 
The lifetime $\tau_{\cal U}$ can be written as
\bea 
\displaystyle\frac {\tau_{\cal U}} {m_j} &=& \displaystyle\frac {16 \pi^2 
d_{\cal U} (d_{\cal U}^2 - 1)} {A_d |\lambda_\nu^{ij}|^2} \left ( \frac {\Lambda_{\cal U}^2} {m_j^2} 
\right)^{d_{\cal U} - 1} \displaystyle\frac {1} {m_j^2}\,\, ,
\label{form5}
\eea
where $m_j$ is the mass of the decaying neutrino.

The normalization constant \cite{georgi1} in the above equation 
(Eq. (\ref{form5})) is defined 
as 
\bea
A_d &=& \displaystyle\frac {16 \pi^{5/2}} {(2\pi)^{2d_{\cal U}}} 
\displaystyle\frac 
{\Gamma (d_{\cal U} + 1/2)} {\Gamma (d_{\cal U} - 1) \Gamma (2d_{\cal U})}\,\, .
\label{form6}
\eea

In the decay process for the four flavour scenario the lightest mass state 
$|\nu_1 \rangle$ is stable, because it does not decay and all other states 
$|\nu_2 \rangle, |\nu_3 \rangle$ and $|\nu_4 \rangle$ are unstable. We can 
state that the total flux of a given energy is negligibly effected by the flux 
of daughter neutrinos having reduced energy and the coherence is lost 
\cite{beacom} (with 
$\Delta m^2 L/E >> 1$ for UHE neutrinos from distant GRB and the oscillatory 
part is absent). The flux for a neutrino $|\nu_\alpha \rangle$ of flavour $\alpha$ on reaching 
the Earth from distant sources like GRB is given as 
\bea
\phi_{\nu_\alpha} (E) &=& \displaystyle\sum_{i} \displaystyle \sum_{\beta} 
\phi_{\nu_\beta}^s |U_{\beta i}|^2 |U_{\alpha i}|^2 exp (-4\pi L/(\lambda_d)_i)\,\, .
\label{form7}
\eea
In Eq. (\ref{form7}) $\alpha, \beta$ indicate the flavour indices and $i$ is 
defined as mass index, $L$ is the baseline length, $U_{\alpha i}$ etc. denote 
the elements of PMNS matrix. For the 4 flavour scenario ( the minimal extension of 3 flavour case by a sterile neutrino) the PMNS matrix can be wriiten as 
\cite{kang}
{\small
\bea
\tilde{U}_{(4 \times 4)} &=& \left (\begin{array}{cccc}
c_{14} & 0 & 0 & s_{14} \\
-s_{14}s_{24} & c_{24} & 0 &c_{14}s_{24} \\
-c_{24}s_{14}s_{34} & -s_{24}s_{34} & c_{34} & c_{14}c_{24}s_{34} \\
-c_{24}s_{14}c_{34} & -s_{24}c_{34} & -s_{34} & c_{14}c_{24}c_{34}
\end{array} \right ) \times
\left (\begin{array}{cccc}
{{U}}_{e1} & {{U}}_{e2} & {{U}}_{e3} & 0 \\
{{U}}_{\mu1} & {{U}}_{\mu2} & {{U}}_{\mu3} & 0 \\
{{U}}_{\tau1} & {{U}}_{\tau2} & {{U}}_{\tau3} & 0 \\
0 & 0 & 0 & 1 \end{array} \right )
\eea
}
{\small
\bea
&=& \left (\begin{array}{cccc}
c_{14}{{U}}_{e1} & c_{14}{{U}}_{e2} & c_{14}{{U}}_{e3} & s_{14}  \\
& & & \\
-s_{14}s_{24}{{U}}_{e1}+c_{24}{{U}}_{\mu1} &
-s_{14}s_{24}{{U}}_{e2}+c_{24}{{U}}_{\mu2} &
-s_{14}s_{24}{{U}}_{e3}+c_{24}{{U}}_{\mu3} & c_{14}s_{24}  \\
&&& \\
\begin{array}{c}
-c_{24}s_{14}s_{34}{{U}}_{e1}\\
-s{24}s{34}{{U}}_{\mu1}\\
+c_{34}{{U}}_{\tau1} \end{array} &
\begin{array}{c}
-c_{24}s_{14}s_{34}{{U}}_{e2}\\
-s{24}s{34}{{U}}_{\mu2}\\
+c_{34}{{U}}_{\tau2} \end{array}  &
\begin{array}{c}
-c_{24}s_{14}s_{34}{{U}}_{e3}\\
-s{24}s{34}{{U}}_{\mu3}\\
+c_{34}{{U}}_{\tau3} \end{array}  &
c_{14}c_{24}s_{34}    \\
&&& \\
\begin{array}{c}
-c_{24}c_{34}s_{14}{{U}}_{e1}\\
-s_{24}c_{34}{{U}}_{\mu1}\\
-s_{34}{{U}}_{\tau1} \end{array}  &
\begin{array}{c}
-c_{24}c_{34}s_{14}{{U}}_{e2}\\
-s_{24}c_{34}{{U}}_{\mu2}\\
-s_{34}{{U}}_{\tau2} \end{array}  &
\begin{array}{c}
-c_{24}c_{34}s_{14}{{U}}_{e3}\\
-s_{24}c_{34}{{U}}_{\mu3}\\
-s_{34}{{U}}_{\tau3} \end{array}  &
c_{!4}c_{24}c_{34}  \end{array} \right )\,\,  ,
\label{form8}
\eea
}
where $U_{\alpha i}$ represents the matrix elements of 3 flavour neutrino mixing matrix $U$, which is given as
\bea
{\cal{U}} &=& \left (\begin{array}{ccc}
c_{12}c_{13} & s_{12}s_{13} & s_{13} \\
-s_{12}c_{23}-c_{12}s_{23}s_{13} & c_{12}c_{23}-s_{12}s_{23}s_{13} &
s_{23}c_{13} \\
s_{12}s_{23}-c_{12}c_{23}s_{13} & -c_{12}s_{23}-s_{12}c_{23}s_{13} &
c_{23}c_{13}  \end{array} \right )\,\,  .
\label{form9}
\eea

In Eq. (\ref{form7}) $\phi_{\nu_\alpha}$ represents the fluxes 
of $\nu_{\alpha}$ 
and $\phi_{\nu_\beta}^s $ is the fluxes of neutrinos having flavour $\beta$ at 
the source. The decay length ($(\lambda_d)_i$) in the Eq. (\ref{form7}) can be 
expressed as 
\bea
(\lambda_d)_i &=& 2.5 {\rm Km} \displaystyle \frac {E} {\rm GeV} 
\displaystyle\frac{\rm {ev}^2} {\alpha_i}\,\, ,
\label{form10}
\eea
where $\alpha_i$ is defined as $m_i/\tau_{\cal U}$, $\tau_{\cal U}$ being the 
neutrino decay 
lifetime. Eq. (\ref{form10}) shows that the decay length ($(\lambda_d)_i$) is a 
function of neutrino energy ($E$).



Applying the equation Eq. (\ref{flux}) and by considering the condition that 
the lightest mass state $|\nu_1 \rangle$ is 
stable we can write the flux of neutrino flavours for four flavour cases on 
reaching the Earth as \cite{athar}-\cite{madhu}
{\small
\bea
\phi^4_{\nu_e} &=& [ {\mid {\tilde{U}}_{e1} \mid}^2 (1 + {\mid {\tilde{U}}_{\mu1} \mid}^2 - {\mid {\tilde{U}}_{\tau1} \mid}^2  - {\mid {\tilde{U}}_{s1} \mid}^2 )\nonumber\\
&&+ {\mid {\tilde{U}}_{e2} \mid}^2 (1 + {\mid {\tilde{U}}_{\mu2} \mid}^2  -
{\mid {\tilde{U}}_{\tau2} \mid}^2  - {\mid {\tilde{U}}_{s2} \mid}^2 ) 
{\rm exp} (-4 \pi L/(\lambda_d)_2)\nonumber \\
& &  + {\mid {\tilde{U}}_{e3} \mid}^2 (1 + {\mid {\tilde{U}}_{\mu3} \mid}^2 -
{\mid {\tilde{U}}_{\tau3} \mid}^2 - {\mid {\tilde{U}}_{s3} \mid}^2 ) 
{\rm exp} (-4 \pi L/(\lambda_d)_3)\nonumber\\
&&+ {\mid {\tilde{U}}_{e4} \mid}^2 (1 + {\mid {\tilde{U}}_{\mu4} \mid}^2 -
{\mid {\tilde{U}}_{\tau4} \mid}^2 - {\mid {\tilde{U}}_{s4} \mid}^2 ) {\rm exp} (-4 \pi L/(\lambda_d)_4)]\phi_{\nu_e}^s\,\,\, ,\nonumber\\
\phi^4_{\nu_\mu} &=& [ {\mid {\tilde{U}}_{\mu1} \mid}^2 (1 + 
{\mid {\tilde{U}}_{\mu1} \mid}^2 - {\mid {\tilde{U}}_{\tau1} \mid}^2  - 
{\mid {\tilde{U}}_{s1} \mid}^2 )\nonumber\\
&&+ {\mid {\tilde{U}}_{\mu2} \mid}^2 (1 + {\mid {\tilde{U}}_{\mu2} \mid}^2  -
{\mid {\tilde{U}}_{\tau2} \mid}^2  - {\mid {\tilde{U}}_{s2} \mid}^2 )
{\rm exp} (-4 \pi L/(\lambda_d)_2)\nonumber\\
& &+ {\mid {\tilde{U}}_{\mu3} \mid}^2 (1 + {\mid {\tilde{U}}_{\mu3} \mid}^2 -
{\mid {\tilde{U}}_{\tau3} \mid}^2 - {\mid {\tilde{U}}_{s3} \mid}^2 )
{\rm exp} (-4 \pi L/(\lambda_d)_3)\nonumber\\
&& + {\mid {\tilde{U}}_{\mu4} \mid}^2 (1 + {\mid {\tilde{U}}_{\mu4} \mid}^2 -
{\mid {\tilde{U}}_{\tau4} \mid}^2 - {\mid {\tilde{U}}_{s4} \mid}^2 )
{\rm exp} (-4 \pi L/(\lambda_d)_4)]\phi_{\nu_e}^s\,\,\, ,\nonumber\\
\phi^4_{\nu_\tau} &=& [ {\mid {\tilde{U}}_{\tau1} \mid}^2 (1 + 
{\mid {\tilde{U}}_{\mu1} \mid}^2 - {\mid {\tilde{U}}_{\tau1} \mid}^2  - 
{\mid {\tilde{U}}_{s1} \mid}^2 )\nonumber\\ 
&&+ {\mid {\tilde{U}}_{\tau2} \mid}^2 (1 + 
{\mid {\tilde{U}}_{\mu2} \mid}^2  - {\mid {\tilde{U}}_{\tau2} \mid}^2  - 
{\mid {\tilde{U}}_{s2} \mid}^2 ){\rm exp} (-4 \pi L/(\lambda_d)_2)\nonumber\\
& & + {\mid {\tilde{U}}_{\tau3} \mid}^2 (1 + {\mid {\tilde{U}}_{\mu3} \mid}^2 -
{\mid {\tilde{U}}_{\tau3} \mid}^2 - {\mid {\tilde{U}}_{s3} \mid}^2 )
{\rm exp} (-4 \pi L/(\lambda_d)_3)\nonumber\\
&&+ {\mid {\tilde{U}}_{\tau4} \mid}^2 (1 + {\mid {\tilde{U}}_{\mu4} \mid}^2 -
{\mid {\tilde{U}}_{\tau4} \mid}^2 - {\mid {\tilde{U}}_{s4} \mid}^2 )
{\rm exp} (-4 \pi L/(\lambda_d)_4)]\phi_{\nu_e}^s\,\,\, ,\nonumber\\
\phi^4_{\nu_s} &=& [ {\mid {\tilde{U}}_{s1} \mid}^2 (1 + {\mid {\tilde{U}}_{\mu1} \mid}^2 - {\mid {\tilde{U}}_{\tau1} \mid}^2  - {\mid {\tilde{U}}_{s1} \mid}^2 )\nonumber\\
&&+ {\mid {\tilde{U}}_{s2} \mid}^2 (1 + {\mid {\tilde{U}}_{\mu2} \mid}^2  -
{\mid {\tilde{U}}_{\tau2} \mid}^2  - {\mid {\tilde{U}}_{s2} \mid}^2 )
{\rm exp} (-4 \pi L/(\lambda_d)_2)\nonumber\\
& & + {\mid {\tilde{U}}_{s3} \mid}^2 (1 + {\mid {\tilde{U}}_{\mu3} \mid}^2 -
{\mid {\tilde{U}}_{\tau3} \mid}^2 - {\mid {\tilde{U}}_{s3} \mid}^2 )
{\rm exp} (-4 \pi L/(\lambda_d)_3)\nonumber\\
&& + {\mid {\tilde{U}}_{s4} \mid}^2 (1 + {\mid {\tilde{U}}_{\mu4} \mid}^2 -
{\mid {\tilde{U}}_{\tau4} \mid}^2 - {\mid {\tilde{U}}_{s4} \mid}^2 ){\rm exp} (-4 \pi L/(\lambda_d)_4)]\phi_{\nu_e}^s\,\,\, .
\label{form12}
\eea
}

%
%
 
In the above both Eqs. (\ref{form12})  $\phi^4_{\nu_\alpha}$  
represents the neutrino fluxes for four 
flavour cases.

In case of $L >> \lambda_d$, Eq. (\ref{form7}) is then reduced to 
\bea
\phi_{\nu_\alpha} (E) &=& \displaystyle\sum_{i(stable), \beta}  
\phi_{\nu_\beta}^s |U_{\beta i}|^2 |U_{\alpha i}|^2 \,\, .
\label{form14}
\eea
Eq. (\ref{form14}) indicates that with the condition $L >> \lambda_d$, the 
decay term is removed because the neutrino decay is completed by the time it 
reaches the Earth. So only the stable state $|\nu_1 \rangle$ exists. So the 
flavour ratio in 4 flavour scenario in this case is changed to 
$|U_{e1}|^2:|U_{\mu 1}|^2:|U_{\tau 1}|^2:|U_{s 1}|^2$ \cite{beacom,pakvasa,farzan}. But when the decay length is close to the baseline length ($\lambda_d \sim L$), then we cannot wash out 
the neutrino decay effect. Therefore the exponential term survives in Eqs. 
(\ref{form12}) and the baseline length ($L$) plays an 
important role. In such cases, considering GRB neutrino fluxes at a fixed 
redshift ($z$) is useful to explore the neutrino decay effects.

\subsection{Detection of UHE neutrinos from a single GRB}

Upward going muons \cite{raj} are produced by the interactions , which are weak 
in nature, 
of $\nu_{\mu}$ or $\bar{\nu}_{\mu}$ with the rock surrounding the Super-K 
detector. While muons from interactions above the detector cannot be sorted out from the continuous rain of muons created in 
cosmic ray showers in the atmosphere above the mountain, muons coming from 
below can only be due to neutrino ($\nu_{\nu}$) charge current interactions 
($\nu_{\mu} + N \rightarrow \mu +X$), since cosmic ray muons cannot make it 
through from the other side of the Earth. Looking upward going muons is the most encouraging way to detect the UHE neutrinos.

The secondary muon yields from the GRB neutrinos can be detected in a detector 
of unit area above a threshold energy $E_{\rm th}$ is given by \cite{nayantara1,raj1,t.k.}
\bea
S  &=&  \int_{E_{\rm th}}^{E_{\nu_{\rm max}}} dE_{\nu} \displaystyle\frac {dN_{\nu}} {dE_{\nu}} P_{\rm shadow}(E_{\nu}) P_{\mu}(E_{\nu}, E_{\rm th})\,\, ,
\label{det8}
\eea
where $P_{\rm shadow} (E_{\nu})$ represents the probability that a neutrino 
reaches the terrestrial detector such as IceCube being unabsorbed by the Earth. We can express this shadow factor in terms of the energy dependent 
neutrino-nucleon 
interaction length $L_{\rm int} (E_{\nu})$ in the Earth and the effective 
path length $X(\theta_z)$ ($\theta_z$ is fixed for a particular single GRB). 
Thus $P_{\rm shadow} (E_{\nu})$ takes the form.
\bea
P_{\rm shadow} = exp[-X (\theta_{z})/ l_{int} (E_{\nu})]\,\, ,
\label{det9}
\eea
where $L_{\rm int} (E_{\nu})$ is given by
\bea
L_{\rm int}(E_{\nu}) = \displaystyle\frac {1} {\sigma_{\rm tot}(E_{\nu}) N_A}\,\, .
\label{det10}
\eea
In the above, $N_A$ is the Avogadro number ($N_A = 6.023 \times 10^23 
{\rm mol}^{-1} = 6.023 \times 10^{23} {\rm cm}^{-3}$) and $\sigma_{\rm tot}$ 
denotes the total cross-section (= charge current cross-section 
($\sigma_{\rm CC}$) + neutral current cross-section ($\sigma_{\rm NC}$)) for 
neutrino absorptions.

The effective path length $X(\theta_z)$ (gm/cm$^2$) can be written as
\bea
X(\theta_z) = \int \rho(r(\theta_z,l)) dl\,\, .
\label{det11}
\eea
We have considered Earth as a spherically symmetric ball having a dense inner 
and outer core and a lower mantle of medium density. So in Eq. (\ref{det11}) 
$\rho(r(\theta_z,l))$ ($l$ is the neutrino path length entering into the Earth) represents the matter density profile inside the Earth, which can be expressed by the Preliminary 
Earth Model (PREM) \cite{prem}.

The probability $P_{\mu} (E_{\nu}, E_{\rm th})$ that a neutrino induced muon 
reaching the detector with an energy above $E_{\rm th}$ can be written as
\bea
P_{\mu}(E_{\nu}, E_{\rm th}) = N_A \sigma_{\rm cc}(E_{\nu})\langle R(E_{\mu}; 
E_{\rm th})\rangle\,\,\, ,
\label{det12}
\eea
where the average muon range in the rock $\langle R(E_{\mu}; E_{\rm th})\rangle $ is given by
\bea
\langle R(E_{\mu}; E_{\rm th})\rangle &=& \displaystyle \frac {1} 
{\sigma_{\rm CC}} \int_{0}^{(1-E_{\rm th}/E_{\nu})} dy R(E_{\nu}(1-y); E_{\rm th}) \times \displaystyle \frac {d\sigma_{\rm CC}(E_{\nu}, y)} {dy} \,\, ,
\label{det13}
\eea
where $y = (E_{\nu} - E_{\mu})/E_{\nu}$ represents the fraction of energy loss 
by a neutrino of energy $E_{\nu}$ in the production of a secondary muons 
having energy $E_{\mu}$. We can replace $E_{\nu} (1-y) $ by $E_{\mu}$ in the 
integrand of Eq. (\ref{det13}). So now the muon range $R(E_{\mu};E_{\rm th})$ 
can be expressed as
\bea
R(E_{\mu}, E_{\rm th}) = \int_{E_{\rm th}}^{E_{\mu}} \displaystyle\frac {dE_{\mu}} {\langle dE_{\mu}/dX \rangle} \simeq \displaystyle\frac{1} 
{\beta} \rm ln \left ( \displaystyle\frac {\alpha + \beta E_{\mu}} 
{\alpha + \beta E_{\rm th}} \right )\,\, .
\label{det14}
\eea
The average energy loss of muon with energy $E_{\mu}$ is given as \cite{t.k.}
\bea
\left \langle \displaystyle\frac{dE_{\mu}} {dX} \right \rangle = - \alpha -
\beta {E_{\mu}} \,\, .
\label{det15}
\eea
The values of the constants $\alpha$ and $\beta$ in Eq. (\ref{det15}), which we have considered in our calculations are
\bea
\alpha &=& {2.033 + 0.077\,\,{\rm ln}[E_{\mu}({\rm GeV})]} \times 10^3\,\,{\rm GeV}\,\,{\rm cm^2}\,\, {\rm gm^{-1}}\,\, ,\nonumber\\
\beta  &=&  {2.033 + 0.077\,\,{\rm ln}[E_{\mu}({\rm GeV})]} \times 10^{-6}\,\,{\rm GeV}\,\,{\rm cm^2}\,\,{\rm gm^{-1}}\,\, ,
\label{det16}
\eea
for $E_{\mu}\,\, \leq \,\, 10^6$  \rm GeV \cite{a.dar} and otherwise 
\cite{guetta}  
\bea
\alpha &=& 2.033\times 10^{-3}\,\, {\rm GeV}\,\, {\rm cm^2}\,\, {\rm gm^{-1}}\, ,\nonumber\\
\beta &=& 3.9 \times 10^{-6}\,\,{\rm GeV}\,\, {\rm cm^2}\,\, {\rm gm^{-1}}\, .
\label{det17}
\eea
In the case of detecting muon events at a 1 Km$^2$ detector such as IceCube 
the flux $\displaystyle \frac{dN_{\nu}} {dE_{\nu}}$ in Eq. (\ref{det8}) is 
replaced by $\phi_{\nu_\mu}^4$ in Eq. (\ref{form12}). 
     
Cosmic tau neutrinos undergo charge current deep inelastic scattering with 
nuclei of the detector material and produces hadronic shower as well as tau 
lepton ($\nu_\tau + N \rightarrow \tau +X$). After traversing some distances, 
which is proportional to the energy of tau lepton, $\tau$ decays into 
$\nu_\tau$ (having diminished energy) and in this process a second hadronic 
shower is induced. These whole double shower processes are introduced as a 
double bang event. The detection of these tau leptons, which are regenerated 
in the lollipop event, is very much complicated due to its noninteracting 
nature with the other particles as they lose energy very fast. The only 
possible way of the detection of tau leptons other than double bang event is 
the production of muons via the decay channel $\nu_\tau \rightarrow \tau 
\rightarrow \nu_{\bar{\mu}}\mu\nu_\tau$ with probability 0.18 \cite{nayan2,tau}. The number 
of such muon events can be computed by solving numerically Eqs 
(\ref{det8} - \ref{det17}) and it is needless to say that  
$\displaystyle \frac{dN_{\nu}} {dE_{\nu}}$ in Eq. (\ref{det8}) is equivalent to $\phi_{\nu_\tau}^4$ (Eq. (\ref{form12})). 

\section{Calculations and Results}
In this section we explore the effect on a flux of neutrinos of different
flavours on reaching the Earth from a distant astrophysical source,
in case such neutrinos undergo unparticle decay along with the usual
mass flavour oscillations. For this purpose we consider a
specific example of ultra high energy neutrinos from
a single GRB and its detection at a kilometer scale Cherenkov detector
such as IceCube. We also assume the existence of a 4th sterile neutrino
in addition to the usual three active flavour neutrinos ($\nu_e$, $\nu_\mu$
and $\nu_\tau$). 

The expression for the
final flux for a neutrino flavour $\alpha$ on reaching the Earth is given
in Eq. (\ref{form7}) along with Eqs. (\ref{form8}-\ref{form12}) (Sect. 2.1). It is to be noted
that the decay part
(${\rm exp}(-4\pi L/(\lambda_d)_i$) for a neutrino mass eigenstate
$|\nu_i \rangle$ will be meaningful and significant for the baseline
length $L \sim (\lambda_d)_i$, the decay length. This decay length depends on
the neutrino-unparticle coupling $\lambda_\nu^{ij}$,
the non-integral scaling dimension $d_{\cal U}$, the dimensional
transmutation scale $\Lambda_{\cal U}$ etc.

The neutrino flux from a single GRB is calculated using
Eqs. (\ref{det1} - \ref{det7}) in
Section 2.1.
We have considered a GRB of energy $E_{\rm GRB} = 10^{53}$ GeV at a redshift
$z = 0.1$ for the present calculations. The measure of distance
(Eq. (\ref{det7}))
corresponding to the chosen redshift is
computed as $10^{15}$ km from the Earth where the values of cosmological
parameters $\Omega_{\Lambda} = 0.68$ and $\Omega_m = 0.3$ are adopted
from PLANCK 2015 data \cite{Planck}. The break energy $E_\nu^{\rm brk}$ is
obtained using Eq. (\ref{det2}) with the value of photon spectrum break energy
$E_{\gamma}^{\rm brk}$ adopted from Table 1 of Ref. \cite{nayantara} for
the Lorentz boost factor $\Gamma = 50.12$. We have considered the current best 
fit values for three neutrino mixing angles ($\theta_{12} = 33.48^{\circ}, 
\theta_{23} = 45^{\circ}$ and $\theta_{13} = 8.5^{\circ}$). The following 
four flavour analysis of different experimental group such as MINOS, Daya Bay, 
Bugey, NOvA \cite{minos1,Adamson:2016jku,bugey,nova1,nova2,nova3,nova4,nova5} 
suggest some limits on four flavour mixing angles ($\theta_{14}, 
\theta_{24}, \theta_{34}$). For $\Delta m_{41}^2 = 0.5\,\, {\rm eV}^2$ NOvA 
\cite{nova2} gives 
the upper limits on $\theta_{24}$ and $\theta_{34}$ as 
$\theta_{24} \leq 20.8^{\circ}$ and $\theta_{34} \leq 31.2^{\circ}$. For the 
same value of $\Delta m_{41}^2$ the upper limits on 
$\theta_{24}$ and $\theta_{34}$ obtained from MINOS \cite{minos1} are 
$\theta_{24} \leq 7.3^{\circ}$ and $\theta_{34} \leq 26.6^{\circ}$. However 
IceCube - DeepCore \cite{icecube} experimental results have proposed 
that 
$\theta_{24} \leq 19.4^{\circ}$ and $\theta_{34} \leq 22.8^{\circ}$ for 
$\Delta m_{41}^2 = 1\,\, {\rm eV}^2$. The limits on $\theta_{14}$ 
are chosen as $1^{\circ} \leq \theta_{14} \leq 4^{\circ}$ in the range
$0.2\,\, {\rm eV}^2 < \Delta m_{41}^2 < 2\,\, {\rm eV}^2$, which is consistent 
with the observational results from the combined experimental analysis by 
MINOS, 
Daya Bay and Bugey-3 \cite{Adamson:2016jku}. By considering the above mentioned limits on four 
flavour mixing angles we have taken $\theta_{14}, \theta_{24} $ and 
$\theta_{34}$ as $3^{\circ}, 5^{\circ}$ and $20^{\circ}$ respectively for our 
calculations. It is to be noted that in the four flavour neutrino decay 
framework the normal hierarchy is evident as we have already discussed in  
Section 2.1.1 that $|\nu_2 \rangle, |\nu_3 \rangle $ and $|\nu_4 \rangle$, 
considering as unstable states, are subjected to undergo unparticle decay 
while only $|\nu_1 \rangle$ is stable. In our calculations we consider $m_2$ 
and $m_3$ as $\sqrt{\Delta m_{32}^2}$ and $\sqrt{2.0 \times \Delta m_{32}^2}$, 
where $\Delta m_{32}^2 = m_3^2 - m_2^2$ 
(normal hierarchy) and $\Delta m_{32}^2 = 2.4 \times 10^{-3}\,\, {\rm eV}^2$ 
(from atmospheric neutrino oscillation) respectively. The value of $m_4$ is estimated from 
$m_4 = \sqrt{\Delta m_{41}^2}$, where $\Delta m_{41}^2$ lies within the 
range $0.2\,\, {\rm eV}^2 < \Delta m_{41}^2 < 2\,\, {\rm eV}^2$. 
By using Eqs. (10 - 22)  we now
calculate the relevant neutrino flux from a single GRB reaching the
detector with or without unparticle decay.

The upgoing secondary muon yield from $\nu_\mu$ in an Earth bound detector
can be computed by using Eqs. (\ref{det8} -
\ref{det17}). We have
considered a kilometer square detector such as IceCube for our present
calculations in case the neutrinos undergo unparticle decay. Note that
we consider UHE neutrinos from a single GRB here. Therefore its directionality
of the neutrino beam with respect to the detector is fixed.

The effect of unparticle decay is characterised mainly by the three parameters
namely, the neutrino-unparticle coupling $\lambda_\nu^{ij}$,
the fractional dimension of
unparticle ($d_{\cal U}$) and the transmutation scale $\Lambda_{\cal U}$. The scale
$\Lambda_{\cal U}$ is fixed at 1 TeV for the present calculations.
The effect of unparticle parameters $d_{\cal U}$ and $\lambda_{\cal U}^{ij}$ are
varied to study how they affect the various quantities that can be measured
at the detector.

In Fig. 1 we plot the variations of neutrino induced muon yields at
the detector for neutrinos from different
single GRBs with different values of GRB energies ($E_{\rm GRB}$).
Since $E_{\rm GRB}$ changes only the normalization constant of single GRB 
neutrino flux
(Eq. 16), the muon yield should increase linearly with the increase
of $E_{\rm GRB}$ as is obtained in Fig. 1 (for a fixed value of
$d_{\cal U}$). In Fig. 1 the dashed line corresponds to the case where the
unparticle decay of neutrinos is considered and the solid line represents the
no decay case.
It is also evident from Fig. 1 that the dashed line (decay case) is
shifted downwards with respect to the solid line (no decay case)
 signifying the depletion of the neutrino flux due to decay.

\begin{figure}[h!]
\centering
 \includegraphics[height=6.0 cm, width=7.5 cm,angle=0]{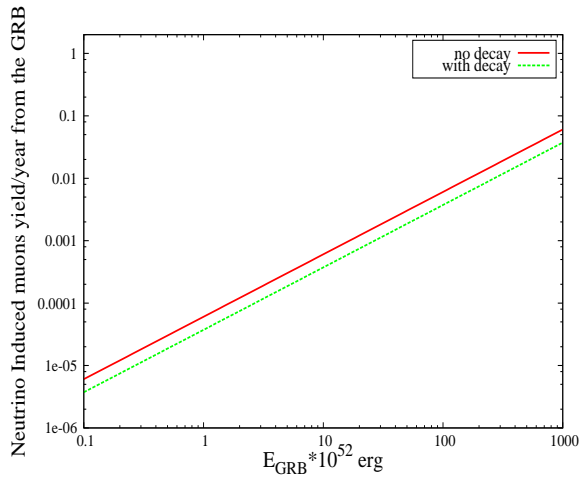} 
\caption{Variation of the neutrino induced muons yield per year  
from the GRB with different energy values of GRB at a fixed zenith angle 
($\theta_z = 160^{\circ}$).}
\label{fig2}
\end{figure}

We show the variations of decay life time of neutrino in terms of 
$\tau/m(=\tau_{\cal U}/m_j)$ for different fixed 
values of $\lambda_\nu^{ij}$ with the unparticle dimension $d_{\cal U}$ in Fig. 2. 
The plots clearly indicate the increasing nature of $\tau/m$ with the increase 
of $d_{\cal U}$, which is manifested in Eq. (14) along with Eq. (15). Fig. 2 
also 
reflects the fact that $\tau/m$ decreases with the reducing values of 
$\lambda_\nu^{ij}$ (Eq. (14)). 

\begin{figure}[h!]
\centering
{\includegraphics[height=6.0 cm, width=7.5 cm,angle=0]{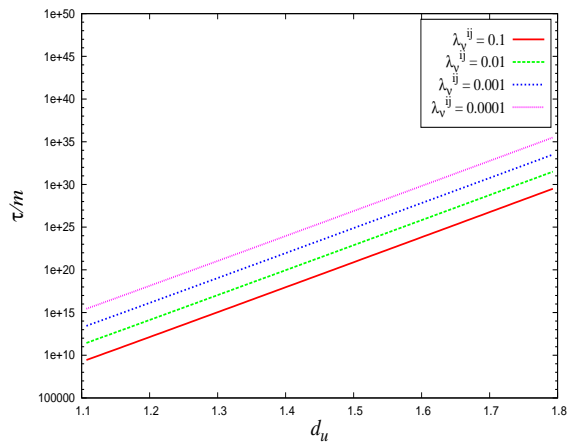}}
\caption{The Variations of the neutrino decay life time ($\tau/m$) 
with the unparticle dimension ($d_{\cal U}$) are shown for four different values 
(0.1, 0.01, 0.001, 0.0001) of couplings $\lambda_\nu^{ij}$. }
\label{fig2}
\end{figure}

Fig. 3 shows the variations of neutrino induced muons at a square kilometer
detector such as IceCube considered here for neutrinos from
different single GRBs at varied redshifts ($z$). We have shown the results
for three fixed values of $\lambda_\nu^{ij}$ as well as for no decay case.
All the plots in Fig. 3 exhibit decrease of neutrino induced muons with
increasing $z$ (the distance of the GRBs from the observer) as is evident from 
Eqs. (6,9).
It is to be noted that the decrease of the coupling $\lambda_\nu^{ij} $
causes the decay length $\lambda_d$  to increase and therefore the depletion
of the neutrino flux (and hence the induced muon yield) will be effective
for neutrinos from GRBs at larger distances or redshifts. For example
in Fig. 3, when $\lambda_\nu^{ij} = 0.0001$ the decay effect is significant
for a GRB with $z \sim 0.1$ whereas for $\lambda_\nu^{ij} = 0.001$ the
depletion due to decay is evident for neutrinos from a nearer
GRB with $z \sim 0.001$.

\begin{figure}[h!]
\centering
{\includegraphics[height=6.0 cm, width=7.5 cm,angle=0]{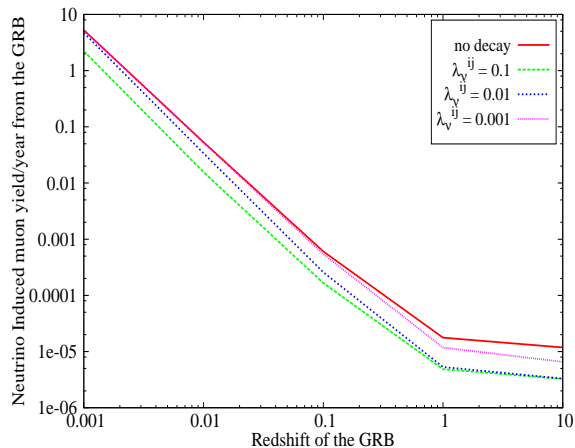}}
\caption{Variations of the neutrino induced muons per year from the GRB 
with different redshifts ($z$) for three different values of $\lambda_\nu^{ij}$ 
as well as for no decay case at a fixed zenith angle 
($\theta_z = 160^{\circ}$). See text for details.}
\label{fig2}
\end{figure}

In Figs. 4(a),4(b) the effects of the unparticle parameters ($d_{\cal U}$ and
$\lambda_\nu^{ij}$) on the unparticle decay of neutrinos are shown.
Comparisons are also made with the cases when only mass-flavour oscillations
are considered. Because of very long baseline the mass flavour oscillations
effect all the neutrino fluxes will be manifested only through an
overall depletion of the flux depending on just the neutrino mixing angles.
The variations of the neutrno induced muon yields at the detector considered
with the unparticle dimension $d_{\cal U}$ for different fixed values of
$\lambda_\nu^{ij}$ are shown in Fig. 4(a). The results with only mass flavour
oscillations (no unparticle decay) are also shown for comparison. All the
calculations are made for UHE neutrinos from a GRB at $z = 0.1$ and at a
zenith angle $\theta_z = 160^{\circ}$. The decay effect is evident in
Fig. 4(a) as the muon yield depletes by $\sim 70$\% from what is expected
for only the mass-flavour case. It can also be noted from Fig. 4(a) that higher
the value of the coupling for unparticle decay of neutrinos, higher is the
unparticle dimension at which the decay effect starts showing up. Since, here
we consider a single GRB at a fixed red shift, the baseline length $L$ is
fixed. Therefore the exponential decay term ${\rm exp}(-L/\lambda_\nu^{ij})$
depends only on the decay length $(\lambda_d)_i$. As the decay length
depends on $\frac {\tau} {m}$ (Eq. (14)) which in turn is a function of
both $d_{\cal U}$ and $\lambda_\nu^{ij}$, the nature of the plots in Fig. 4(a) 
varies accordingly. Similar trends can also be seen when the neutrino induced 
muons are plotted with $\lambda_\nu^{ij}$ for different fixed values of $d_{\cal U}$
(Fig. 4(b)).
\begin{figure}[h!]
\centering
\subfigure[]{
\includegraphics[height=6.0 cm, width=6.0 cm,angle=0]{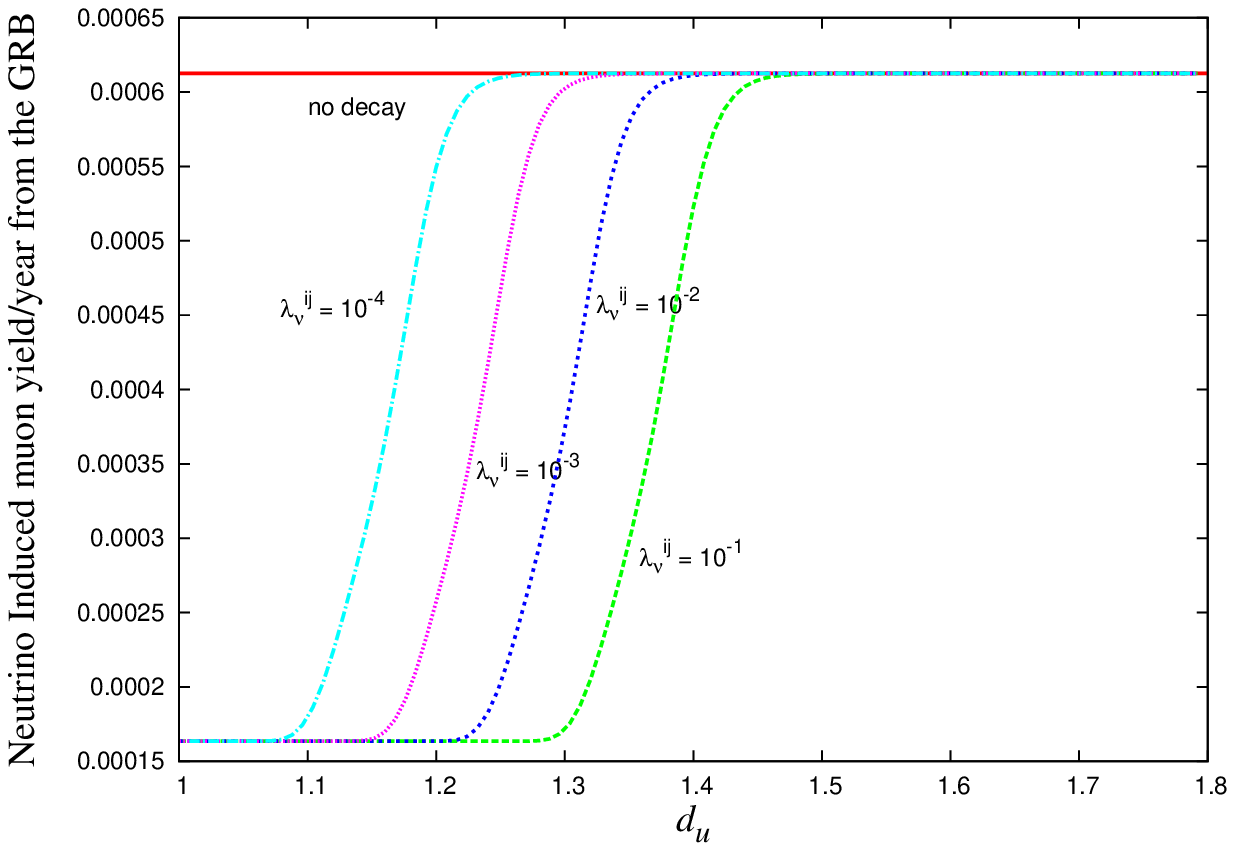}}
\subfigure []{
\includegraphics[height=6.0 cm, width=6.0 cm,angle=0]{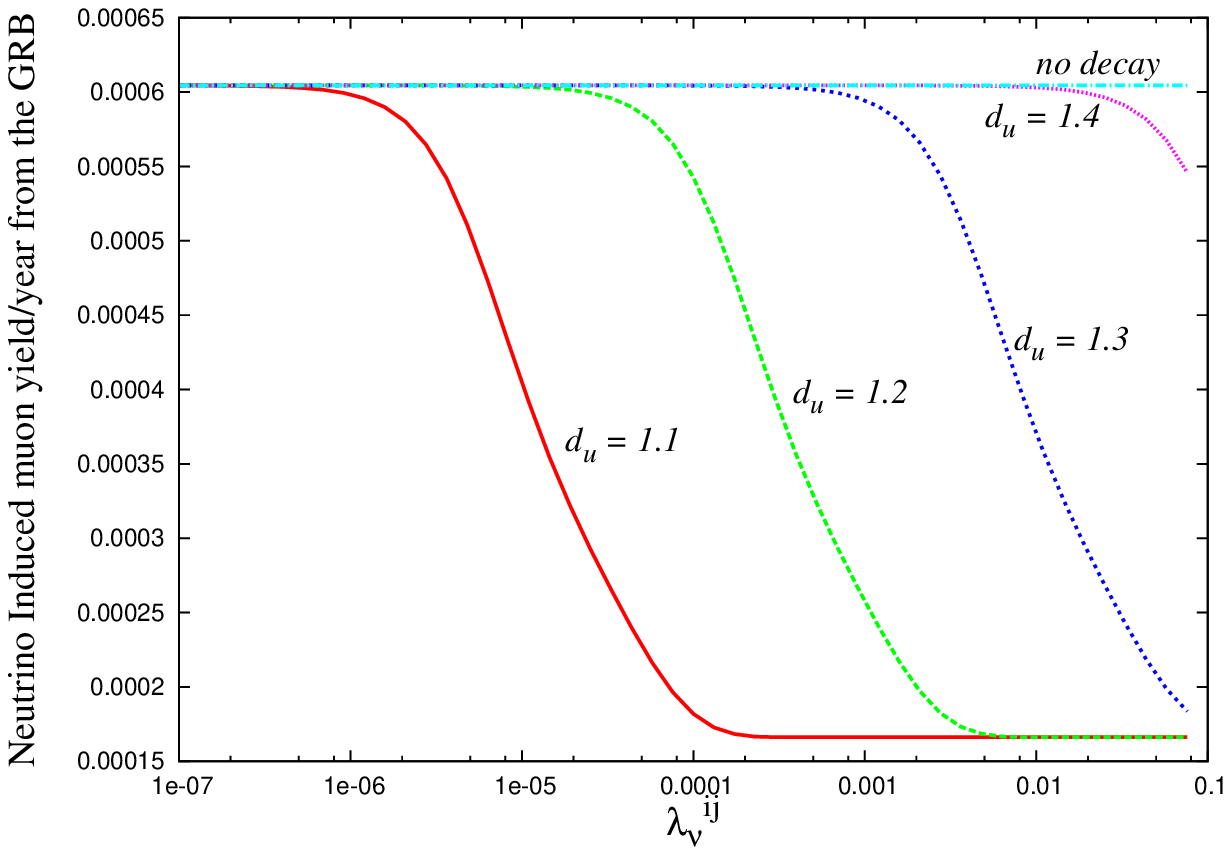}}
\caption{The variations of the neutrino induced upward going muons per year 
from the GRB with (a) different values of $d_{\cal U}$ for four different fixed 
values of $\lambda_\nu^{ij}$ as well as for the mass flavour case 
(no decay case), (b) different values of 
$\lambda_\nu^{ij}$ for four different fixed values of the unparticle dimension 
$d_{\cal U}$ (1.1, 1.2, 1.3, 1.4) and in addition for no decay case. 
See text for details.}
\label{fig1}
\end{figure}
\newpage
\section{Summary and Discussions}

In this work we have explored the possibility of unparticle decay of Ultra 
High Energy (UHE) neutrinos from a distant single GRB and its consequences on 
the neutrino induced muon yields at a kilometer square detector. The concept 
of unparticles first proposed by Georgi from the consideration of the 
presence of a yet unseen scale invariant sector which may be present in the four dimensions with non-renormalizable interactions with Standad Model particles. 
The ``particles'' in this scale invariant sector are termed as ``unparticles''. The unparticle scenario and its interaction with SM particles such as neutrinos are expressed by an effective lagrangian, which is expressed in terms of the 
effective couplings ($\lambda_\nu^{\alpha \beta}$, where $\alpha, \beta$ are 
the flavour indices) between neutrinos ($\nu_{\alpha, \beta}$) and the 
scalar unparticle 
operator (${\cal O}_{\cal U}$), the scaling dimension ($d_{\cal U}$) and the 
dimension transmutaion scale ($\Lambda_{\cal U}$). In the case of the neutrino 
unparticle interaction, heavier neutrinos become unstable and can decay into 
the unparticles and lighter neutrinos. In the present work in order to explore 
the unparticle decay process 
we have considered the UHE neutrino signatures obtained from GRB events for a 
3+1 neutrino frameowrk. We estimate how the effect of an unparticle decay of 
neutrinos in addition to the mass-flavour oscillations can change the 
secondary muon yields from GRB neutrinos at a 1 Km$^2$ detector such as IceCube 
for a four flavour scenario. The advantage of choosing UHE neutrinos from 
GRB is that the oscillatory part is averaged out due to their astronomical 
baslines ($\Delta m^2 L/ E >> 1$). In the present work we consider the neutrino fluxes from a point like source such as a single GRB. We calculate the muon 
yield in such a scenario where both unparticle decay and flavour oscillation 
(suppression) is considered. we also investigate the effect of fractional 
unparticle dimension $d_{\cal U}$ as also the coupling $\lambda_\nu^{ij}$ on the muon 
yield and compare them with the case where only flavour suppression (without 
an unparticle deacy) is considered. It is observed that the effect of unparticle decay considerably affects the muon yield. This is a representative 
calculation to demonstrate the unparticle decay neutrinos can indeed affect 
the neutrino flux from distant sources such as GRBs. But there could be various sources of errors not only in detection processes but also in estimating 
theoretical GRB flux, the neutrino propagation through Earth before being 
detected at the detector. The other experimental uncertainties include the errors that may creep in from Digital Optical Modules (DOMs) that would record 
the muon track events (and shower events). The optical properties of the ice 
such as absorption coefficients and optical scattering and the systematic 
uncertainty associated with it affect the signals at DOM.

{\bf Acknowledgments} : MP thanks the DST-INSPIRE fellowship grant by DST, 
Govt. of India.

\end{document}